\documentclass[prl,aps,tightenlines,floatfix,nofootinbib, twocolumn]{revtex4}

\usepackage{epsf}
\usepackage{graphicx}

\newcommand{\be}{\begin{equation}}
\newcommand{\ee}{\end{equation}}

\begin{document}

\title{Are neutron stars with crystalline colour superconducting cores
interesting for LIGO?}

\author{B. Haskell, N. Andersson, D.I. Jones and L. Samuelsson}
\affiliation{School of Mathematics, University of Southampton,
  Southampton SO17 1BJ, UK}

\begin{abstract}
  We estimate the maximal deformation that can be sustained by a
  rotating neutron star with a crystalline colour superconducting
  quark core. Our results suggest that current gravitational-wave data
  from LIGO have already reached the level where a detection would
  have been possible over a wide range of the poorly constrained QCD
  parameters.  This leads to the non-trivial conclusion that compact
  objects \emph{do not} contain maximally strained colour crystalline
  cores drawn from this range of parameter space.  We discuss the
  uncertainties associated with our simple model and how it can be
  improved in the future.
\end{abstract} 
  
\maketitle

{\em Introduction}. --- A spinning neutron star may be an interesting
gravitational-wave source provided that it is deformed in some way.
Such asymmetries, colloquially referred to as ``mountains'', may arise
in a number of different ways. The elastic crust of the star may
sustain shear stresses \cite{ucb,bryn}, the size of which are limited
by the breaking strain of the crustal lattice. A magnetic star in
which the magnetic field is misaligned visa-vi the rotation axis may
also be deformed in an interesting way \cite{magpaper,wasserman}.
Estimates of the possible level of neutron star asymmetry are of great
current interest given the improving upper limits obtained by LIGO
\cite{LIGO}. The best current limit set by the detectors corresponds
to a deformation of $\epsilon< 7.1\times10^{-7}$ in PSR 2124-3358, a
neutron star spinning at 202.8~Hz. This should be compared to the
maximal mountain size predicted from theory. We have recently
concluded that the largest crust mountain one should expect on an
isolated neutron star is $\epsilon\approx 2.4\times10^{-6}$
\cite{bryn}. This estimate is obtained assuming that the breaking
strain of the crust is $\bar{\sigma}_\mathrm{br} = 10^{-2}$, which may
be optimistic. The comparison shows that the sensitivity of the
detectors have reached the level where the observations are beginning
to confront the theory. Of course, we are still some way away from
testing more conservative models. Recall, for example, that the crust
breaking strain is usually assumed to lie in the range
$\bar{\sigma}_\mathrm{br} = 10^{-5}-10^{-2}$. If the true value is at
the lower end of this range, then we would still be far away from a
detection. It is also important to keep in mind that the theoretical
estimates are based on a maximisation argument. It is not at all clear
that there are physical avenues that lead to a star being deformed at
this level. As a final caveat, we note that there also exist spin-down upper 
limits on ellipticities, where conservation of energy is used to produce
 a moment of inertia and distance dependent bound on $\epsilon$. However, 
these are less robust than the direct upper limits, and will not be 
our focus here.
 
Despite the various caveats, the existing observational data are very
interesting.  As shown by Owen \cite{owen}, maximally strained solid
quark stars could radiate at a level where a detection would have been
possible.  It may also be the case that strong emission could occur
from an elastic phase in the neutron star core.  Core deformations are, in fact,
likely to be more significant than crustal ones, since the high
density region provides a larger contribution to the quadrupole moment.  
The problem has been the lack of ``quantitative''
models. This may have changed recently, with the suggestion of
crystalline colour superconducting phases in the deep core (see
\cite{crystal1,crystal2,krishna} and references therein).  Our aim
with this letter is to adapt our recently developed framework for
estimating the maximal crustal mountains in a neutron star \cite{bryn}
to consider elastic cores. This leads to a simple model for a neutron
star with a crystalline core of deconfined quarks and a normal hadron
fluid envelope. We compare the attainable mountains within this model
to the current limits set by LIGO and ask to what extent observations
are already confronting QCD modelling.

{\em The elastic core model}. --- As discussed in
\cite{crystal1,crystal2} it may be energetically favourable for colour
superconducting quark matter to form a crystalline structure.  The
shear modulus can be estimated as \cite{krishna}
\begin{equation}
\mu=3.96\times 10^{33}\, \mbox{erg/cm}^3\left(\frac{\Delta}{10\mbox{MeV}}\right)^2\left(\frac{\mu_c}{400 \mbox{MeV}}\right)^2
\ .
\end{equation}
It depends on the gap parameter $\Delta$ and  the quark chemical
potential $\mu_c$, for which the authors of \cite{krishna} suggest  estimated ranges
of
\begin{equation}
350\mbox{MeV}< \mu_c < 500\mbox{MeV} \ , \quad 5\mbox{MeV}< \Delta < 25\mbox{MeV} \ .
\label{range}\end{equation}
There is naturally a significant uncertainty associated with these
estimates.  In order to keep the analysis simple, we will consider the
shear modulus to be constant in the core and dependent only on the
combination $(\Delta\mu_c)^2$, a parameter on which one may then be
able to place constraints from observations. In reality,
the various parameters are expected to be (weakly) density dependent,
but it does not seem relevant to try to account for this in a first
study.

\begin{figure*}[t]
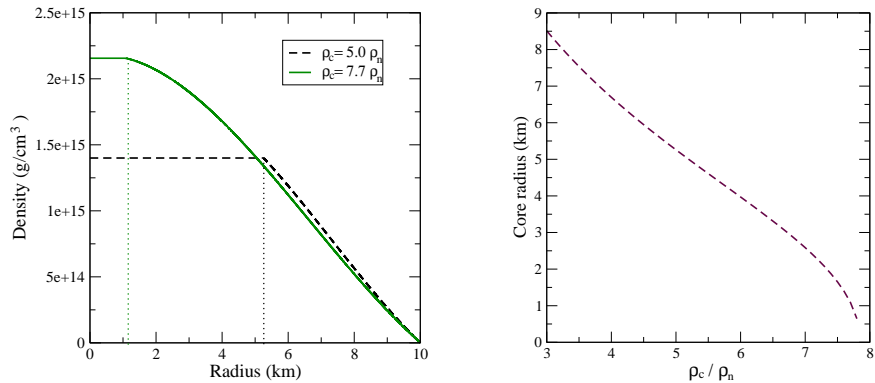

\begin{minipage}[t]{16cm}
\begin{centerline}
{\includegraphics[height=5cm,clip]{densityn.eps}\hspace*{1cm}
\includegraphics[height=5cm,clip]{radiusn.eps}}
\end{centerline}
\caption{The left panel shows the density profiles for two different 
values of the ratio $\rho_c/\rho_\mathrm{n}$. One can see quite clearly
that  (for a fixed mass and radius of
$1.4M_\odot$ and 10~km, respectively) the core becomes significantly smaller 
as the transition density rises. This
fact is also illustrated in the right panel, which shows the radius of
the core as a function of the transition density.}
\label{densplot}
\end{minipage}
\end{figure*}

We  consider a simple model of a neutron star with an
elastic core of deconfined quarks and a fluid exterior. The equation
of state for the exterior is taken to be an $n=1$ polytrope, while the
core is described by an incompressible fluid. This is a reasonable first
approximation for a compact star with a deconfined quark
core. The particular advantage of this simple core-mantle
model is that the Newtonian hydrostatic equilibrium equations
\be
\nabla_a p =-\rho\nabla_a \Phi \ ,
\ee
where $p$ is the pressure, $\Phi$ the gravitational potential and
$\rho$ the density, can be solved analytically. We match the solutions
for the two regions at a given transition density $\rho_c$ and impose
the constraints that the total mass of the star is $M=1.4M_{\odot}$
and the radius is $R=10$ km. These constraints are obviously not
necessary, but they restrict the available parameter space.  This way
one can construct a sequence of stellar models by varying the
transition density, which we take to be in the range $3 <
{\rho}_c/\rho_\mathrm{n} < 8$ (where $\rho_\mathrm{n}$ is the
nuclear saturation density). Two examples of the resulting density
distributions are shown in Figure~\ref{densplot} together with the
radius of the solid core as a function of the transition density.

In order to calculate the maximum ellipticity we extend the formalism
developed in \cite{bryn}. In essence, we perturb the background
configuration assuming that the perturbations have a $Y_{22}(\theta, \varphi)$ angular
dependence. Deviations from sphericity build up strain in the
core. It breaks when $\bar{\sigma}=\bar{\sigma}_\mathrm{br}$ at some
point.  Here, $\bar{\sigma}$ is the modulus of the strain
tensor as defined in \cite{bryn}. As we have already mentioned, the
breaking strain $\bar{\sigma}_\mathrm{br}$ is highly uncertain. To
make progress we will assume that it lies in the usual range
considered for neutron stars, i.e. we take $10^{-5}<
\bar{\sigma}_\mathrm{br}<10^{-2}$.

To maximise the ellipticity we thus assume the maximum strain in the
core and solve the hydrostatic equilibrium equations, which now
include elastic terms due to the deformations:
\begin{equation}
\nabla^a{t_{ab}}=\rho\nabla_{b}\Phi+\nabla_{b} p \ ,
\end{equation}
where we have defined the stress tensor
\begin{equation}
t_{ab}=\mu\left(\nabla_{a}\xi_b+\nabla_b\xi_a-\frac{2}{3}\delta_{ab}\nabla^c\xi_c\right) \ , 
\end{equation}
with $\mu$ the shear modulus, which we consider constant, and $\xi^a$
the components of the displacement vector for the crystalline phase.
As boundary conditions we require regularity at the centre of the star
and continuity of the components of the traction both at the surface and at the
interface between the solid and the fluid, where we also impose that
the deformed shape of the star maximise the strain in the core, as
described in \cite{bryn}.

Having solved for the equilibrium shape of the perturbed star we
can calculate the ellipticity
\begin{equation}
\epsilon=\frac{I_{xx}-I_{yy}}{I_{zz}} \ ,
\label{epsdef}\end{equation}
which, for an observer on the
$z$ axis, is tied to the gravitational-wave strain amplitude by
\be
h =\frac{16\pi^2 G}{c^4}\frac{\epsilon \, I_{zz}\nu^2}{r} \ ,
\label{gwamp}\ee 
where $\nu$ is the neutron star spin frequency, $I_{zz}$ it's
principal moment of inertia and $r$ the distance from Earth
\cite{LIGO}.  In order to facilitate a direction comparison with the
ellipticities discussed in \cite{LIGO}, we fix $I_{zz}=10^{45}$g
cm$^2$ rather than calculating it for each of our stellar models. 
This is just a convention, since it is clear from (\ref{epsdef}) and
(\ref{gwamp}) that $I_{zz}$ actually does not affect the final expression for
the gravitational-wave amplitude.

{\em Results and discussion}. --- The ellipticities that we calculate
depend on the breaking strain $\bar{\sigma}_\mathrm{br}$ and the 
combination $(\Delta \mu_c)^2$. 
The results also depend on the chosen value of
the transition density between the core and the crust. This is
obvious, as one would expect a more significant elastic core to be able
to sustain larger deformations.

\begin{figure*}[t]
\begin{minipage}[t]{16cm}
\begin{centerline}
{\includegraphics[height=5cm,clip]{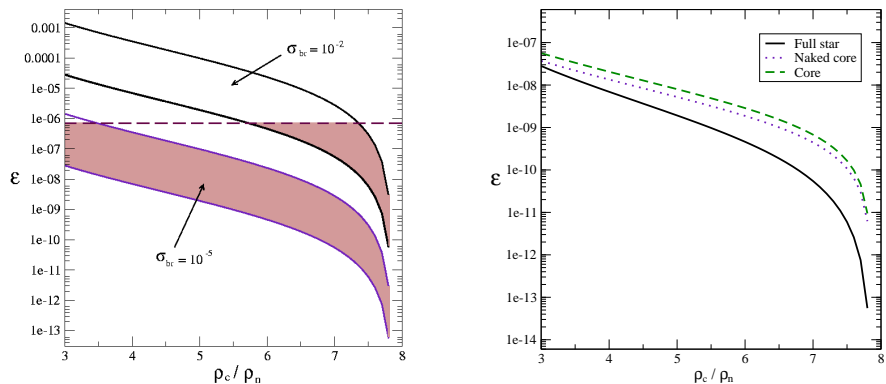}\hspace*{1cm}
 \includegraphics[height=5cm,clip]{coren.eps}}
\end{centerline}
\caption{In the left panel, the ellipticities obtained from our
  analysis are shown as a function of the core transition density
  $\rho_c/\rho_\mathrm{n}$, where $\rho_\mathrm{n}$ is the nuclear
  saturation density.  We consider two breaking strains, a maximum
  value $\bar{\sigma}_\mathrm{br}=10^{-2}$ and a minimum of
  $\bar{\sigma}_\mathrm{br}=10^{-5}$. For each of these values we
  examine the region between the maximum and the minimum shear modulus
  $\mu$, obtained for the range of parameters in \cite{krishna}, cf.
  (\ref{range}). The shaded region indicates the region permitted by
  LIGO observations and the horizontal line is the current LIGO upper
  limit of $\epsilon=7.1 \times 10^{-7}$ for PSR J2124-3358
  \cite{LIGO}.  If the breaking strain is close to the maximum the
  observations are already confronting the theory, showing that
  J2124-3358 at least does not contain such a maximally strained core.
  If the breaking strain is close to the minimum, however, no such
  conclusion can be drawn.  In the right panel we compare the
  ellipticity obtained by considering the full star, only the core of
  the full star and just the naked core, with no fluid around it. We
  take the breaking strain to be $\bar{\sigma}_\mathrm{br}=10^{-5}$.
  As one can see the results for the two cores do not differ
  significantly, while the ellipticity of the core plus fluid star is
  smaller, more so as the core size decreases at higher transition
  densities.}
\label{results}
\end{minipage}
\end{figure*}

Our main results are illustrated in the left panel of Figure
\ref{results}, where we show the obtained ellipticities for both the
maximum breaking strain usually considered for a neutron star,
$\bar{\sigma}_\mathrm{br}=10^{-2}$, and a more conservative value,
$\bar{\sigma}_\mathrm{br}=10^{-5}$. For each of these values we
consider the region between the maximum and the minimum value of the
shear modulus $\mu$ for the range of parameters suggested in
\cite{krishna}, cf. (\ref{range}). We compare these ellipticities to
the best current upper limit set by LIGO using the S3/S4 science runs,
which is $\epsilon=7.1\times 10^{-7}$ for PSR J2124-3358 \cite{LIGO}.
It is quite clear that, if the maximum breaking strain applies and the
star is maximally deformed, then LIGO would have made a detection. Such 
a detection would have implied, cf. Figure~\ref{results}, that the transition to a quark core would
 have to take place below $\rho\approx 7.5 \rho_\mathrm{n}$.  The fact that no 
detection was made rules out such a
scenario.  Of course, this non-detection does not say whether it is
the QCD model that is at fault (i.e. there is no such high density
core) or whether the QCD core exists but simply isn't significantly
strained.

These results are obviously interesting, especially since the
sensitivity of gravitational wave searches are set to improve.
Looking ahead, the soon to be completed S5 science run will provide
approximately one year's worth of data at initial design sensitivity.
The S3/S4 data spanned approximately one month with a noise floor at
near twice design sensitivity, so an upper bound from the S5 run will
give a maximum ellipticity a factor of $2 \cdot \sqrt{12}$ tighter
than for S3/S4 (the upper bound scales linearly with the noise floor
and as the inverse square root of the observation duration).  For
pulsar J2124-3358 this would correspond to a maximum ellipticity of
approximately $1 \times 10^{-7}$.  This could allow
detections even if the breaking strain were at the lower end of the
range we have considered.  For instance, if
$\bar{\sigma}_\mathrm{br}=10^{-5}$ then detection would be possible if
the crust-core transition occurs at a density no greater than
$\rho\approx 5 \rho_\mathrm{n}$, assuming parameters at the upper end
of the range given in (\ref{range}).  Looking further ahead, one year
of data from Advanced LIGO would yield a further improvement of a
factor of ten in sensitivity, corresponding to $\epsilon = 1 \times
10^{-8}$ for J2124-3358.  A detection would then be possible providing
the crust-core transition density lies below $\rho\approx 3
\rho_\mathrm{n}$ \emph{regardless of where the shear modulus and
  breaking strain parameters lie in their respective likely ranges}. A
non-detection (i.e. upper bound) would demonstrate that the star in
question \emph{does not} contain such a maximally strained core with
the appropriate transition density.

Our analysis suggests neutron stars containing colour crystalline
cores are promising candidates for gravitational wave detection, and that
in the event of no detection being made the corresponding upper limits
place non-trivial constrains on the deep interiors of compact objects,
showing that either the strain in the core is small or that the high
density equation of state does not support such solid phases.  Of
course, the model we have considered is very simplistic, and needs to
be improved in a number of ways if we want the results to be truly
quantitative.  Most important would be to improve our understanding of
the crystalline colour superconductor. We have assumed that it behaves
like ordinary elastic matter with a given shear modulus.  This is a
natural first assumption, but it may not be the complete picture.  For
example, we have not accounted for the ``superfluid'' nature of the
core. Yet this could be an important omission.  Perhaps a comparison
between the crystalline colour superconductor and supersolid Helium
\cite{kim,dorsey,joss} would help improve our understanding? This is
an interesting possibility since supersolid Helium is amenable to
laboratory experiments.

Up to this point we have assumed that the nature of the phase
transition at the interface between the quark core and the hadron
fluid is such that there is no discontinuity in the density profile.
However,
there could be a (potentially sizeable) discontinuity at the
interface, see for example Fig. 4 in \cite{alford}.  To gain some
insight into how this might affect our results, we have carried out a
comparison of the ellipticities of three stars that can be easily
modelled within our framework.  Firstly we considered a star
consisting of the usual solid core surrounded by fluid, calculating
$I_{xx}-I_{yy}$ (and hence the ellipticity) by integrating over the
whole star in the conventional manner.  Then we took the same star but
integrated over the core only.  Finally we took a bare core with no
fluid and with the same mass and radius as the core in the first two
models.  The results of these calculations (as a function of core
density) are shown in the right panel of Figure \ref{results}.

Interestingly, there is little difference between the quadrupole of
the bare core and the quadrupole integrated over only the core of the
core plus fluid star.  However, the \emph{addition} of the fluid
envelope serves to \emph{decrease} the ellipticity, the decrease being
more severe for stars with higher core-fluid transition densities.
This decrease can be understood in a simple intuitive way.  Unlike the
solid, the fluid cannot support shear strains, so its perturbation
from spherical symmetry is sourced entirely by the perturbed
gravitational potential.  The fluid's outer surface will therefore be
close to spherical.  The fluid therefore serves to `fill in' the
depressions of the solid core, resulting in a partial cancellation
between the $I_{xx}-I_{yy}$ contributed by the core and fluid.
Clearly, this cancellation will be more substantial the closer the
density of the core and fluid, the case of a smooth density transition
and a bare core being extreme cases.

This leads us to conjecture that in the case of a substantial density
step at the core-envelope interface, the cancellation would be far from
complete, leading to an ellipticity closer to the bare core values
shown in the right panel of figure \ref{results}.  For a small core, 
this could lead to an order of magnitude increase in the ellipticities
calculated assuming a smooth density transition, increasing further
the importance of gravitational-wave observations in confronting QCD.
A rigorous treatment of such a star should employ a realistic equation
of state (such as that employed in \cite{alford}) which would
naturally incorporate any density step. 

The model we have considered is obviously simplistic in many other
ways.  In particular, one would want to relax the assumption that
$\Delta \mu_c$ is constant. In order to make real improvements, one
would (again) want to use a realistic equation of state.
This would require the calculation to be carried out within General
Relativity.  Then the determination of the background solution is
straightforward, but the calculation of the mountain size would
require implementing the General Relativistic theory of elasticity,
see for example \cite{lars}.  To date, there have been no such
calculations.  Work in this direction should clearly be encouraged.

Finally, we need to improve our understanding of the breaking strain.
The range of values that we have used, $10^{-5} \le
\bar{\sigma}_\mathrm{br} \le 10^{-2}$, is relevant for a crust
consisting of normal matter. However, the physics of
the core is very different from that of the crust. 
There is no reason to believe that the
estimates on the breaking strain for the crust should be applicable to
the core. 
The response of the crystalline quark matter
to large stresses is also uncertain. Normal matter will be
predominantly brittle and break into pieces when the temperature is
sufficiently far below the melting temperature and will respond by
plastic flow (up to some limit) otherwise. How an elastic quark core
will respond is completely unknown.  Yet, for our purposes it may not
matter which scenario is realised as long as the timescale for plastic
flow at a given strain is longer than the observation time.

\acknowledgments
We thank Krishna Rajagopal for useful discussions. This work was supported by 
PPARC/STFC via grant numbers PP/E001025/1 and PP/C505791/1.


\begin{thebibliography}{10}

\bibitem{ucb} G. Ushomirsky, C. Cutler, L. Bildsten, MNRAS {\bf 319} 902 (2000)

\bibitem{bryn} B. Haskell, D.I. Jones, N. Andersson, MNRAS {\bf 373} 1423 (2006)

\bibitem{magpaper} B. Haskell, L. Samuelsson, K. Glampedakis, N. Andersson, 
{\em Modelling magnetically deformed neutron stars}, preprint
arXiv:0705.1780

\bibitem{wasserman} T. Akgun, I. Wasserman, {\em Toroidal Magnetic Fields in Type II Superconducting Neutron Stars}, preprint arXiv:0705.2195

\bibitem{LIGO} B. Abbott et al, {\em Upper limits on gravitational wave emission from 78 radio pulsars}, preprint arXiv:gr-qc/0702039

\bibitem{owen} B.J. Owen, Phys. Rev. Lett. {\bf 95} 211101 (2005) 


\bibitem{crystal1}  K. Rajagopal, R. Sharma, Phys. Rev. D. {\bf 74} 094019 (2006)

\bibitem{crystal2} K. Rajagopal, R. Sharma, J. Phys. G. {\bf 32} S483 (2006)

\bibitem{krishna} M. Mannarelli, K. Rajagopal, R. Sharma, {\em The rigidity of crystalline color superconducting quark matter}, preprint arXiv:hep-ph/0702021

\bibitem{kim} E. Kim, M.H.W. Chan, Science {\bf 305} 1951 (2004)

\bibitem{dorsey} A.T. Dorsey, P.M. Goldbart, J. Toner, Phys. Rev. Lett. {\bf 96} 055301 (2006) 

\bibitem{joss} C. Josserand, T. Pomeau, S. Rica, Phys. Rev. Lett. {\bf 98} 
195301 (2007)  

\bibitem{alford} M. Alford, S. Reddy, {\em Compact stars with color superconducting quark matter}, preprint nucl-th/0211046

\bibitem{lars} M. Karlovini, L. Samuelsson, Class. Quantum Grav. 
{\bf 20}  3613 (2003)

\end{thebibliography}
\end{document}